\documentclass[aps,prb,twocolumn,superscriptaddress, longbibliography,floatfix]{revtex4-2}
\usepackage{graphicx}
\usepackage{latexsym}
\usepackage{amssymb}
\usepackage{amsmath}
\usepackage{amsfonts}
\usepackage[vcentermath]{youngtab}
\usepackage{upgreek}
\usepackage{bm,dsfont}
\usepackage{multirow}
\usepackage{enumitem}
\usepackage[usenames, dvipsnames]{color}
\usepackage[svgnames]{xcolor}
\usepackage{comment}
\usepackage{hyperref}

\hypersetup{
colorlinks = true,
linkcolor = [rgb]{0.70,0.13,0.13}, 
citecolor = [rgb]{0.13,0.55,0.13},
urlcolor = [rgb]{0.25, 0.41, 0.88}}
\usepackage{makecell}
\usepackage{dcolumn}
\usepackage{array}
\newcolumntype{d}{D{.}{.}{-1}}

\newcommand{\up}{\uparrow}
\newcommand{\dn}{\downarrow}
\newcommand{\pd}{\partial}
\newcommand{\e}{\mathrm{e}}

\newcommand{\la}{\langle}
\newcommand{\ra}{\rangle}
\newcommand{\Y}{\mathrm{Y}}
\newcommand{\Yp}{\mathrm{Y}'}
\newcommand{\Jt}{\mathcal{J}}

\begin{document}

\title{Spin supersolid in triangular lattice easy-axis spin-1/2 antiferromagnet: a fermionic parton perspective}

\author{Meng Zeng} 
\affiliation{Max Planck Institute for the Physics of Complex Systems, Nöthnitzer Strasse 38, 01187 Dresden, Germany}
\author{Roderich Moessner} 
\affiliation{Max Planck Institute for the Physics of Complex Systems, Nöthnitzer Strasse 38, 01187 Dresden, Germany}
\author{Shu Zhang} 
\affiliation{Collective Dynamics and Quantum Transport Unit, Okinawa Institute of Science and Technology Graduate University, Onna-son 904-0495, Japan}
\affiliation{Institute for Theoretical Solid State Physics, IFW Dresden, Helmholtzstrasse 20, 01069 Dresden, Germany}
\affiliation{Max Planck Institute for the Physics of Complex Systems, Nöthnitzer Strasse 38, 01187 Dresden, Germany}

\date{\today}

\begin{abstract}
We study the supersolid order of the spin-$\tfrac12$ easy-axis XXZ antiferromagnet on the triangular lattice using a fermionic
parton mean-field theory built on the $\pi$-flux Dirac spin liquid, with the mean-field order generated by an on-site spinon interaction term and nearest-neighbor (anisotropic) spin couplings. The collective modes are obtained using the random-phase approximation (RPA). In a finite parameter region, we find the $\Y$ state as a self-consistent solution, together with the competing
$\Yp$ state, which is a $90^\circ$-rotated $\Y$ state inside the order plane. The mean-field energy favors
$\Yp$, even though the two states are exactly degenerate in the classical limit. 
However, the $\Y$ state can become lower in energy when higher-order corrections due to the fluctuations of the collective modes are considered. The $\Y$ state can also be stabilized on the mean-field level by applying small external fields, which we do in two different ways: moment-aligned local field and uniform field along the easy axis. 
In both cases, the RPA calculation for the stabilized $\Y$ state gives a roton minimum at the $M$ point in the excitation spectrum. A gapped pseudo-Goldstone mode above the gapless Goldstone mode is obtained as well, which has also been observed experimentally.
Furthermore, we investigate the effects of tuning the spinon interaction strength and the anisotropy in spin couplings in our phenomenological model. Comparison with the recent numerical and experimental results is also discussed.

\end{abstract}

\maketitle


\section{Introduction}
\label{sec:intro}

A supersolid breaks lattice translations and a continuous $U(1)$ symmetry at the same time. One of the
cleanest lattice realization is the spin-$\tfrac12$ antiferromagnet on the triangular lattice with
easy-axis exchange anisotropy. Mapped onto hard-core bosons, $S^z$ becomes the boson density and
$S^\pm$ create and remove bosons. In this language the model describes bosons with nearest-neighbor
repulsion and hopping, and its ground state is a supersolid~\cite{Wessel2005,Heidarian2005,Melko2005,Wang2009,Jiang2009,Heidarian2010,Yamamoto2014,Sellmann2015}. In spin
language it is the $\Y$ state, where the magnetic orders on each triangular plaquette form a coplanar $\Y$-shaped pattern. The $S^z$ components form a three-sublattice density wave, and
in-plane order breaks the $U(1)$ symmetry about the easy axis. Classically, $\Y$ is degenerate with
other coplanar states, among them the $\Yp$ state, a $90^\circ$-rotated $\Y$ state inside the order plane~\cite{Gallegos2024}, and quantum fluctuations select $\Y$ as the true ground state~\cite{Miyashita1986,Kleine1992,ShengHenley1992,Gallegos2024}.

Cobalt-based triangular magnets have made the model experimentally accessible. Na$_2$BaCo(PO$_4$)$_2$~\cite{Sheng2022,Gao2022,Sheng2025} and
K$_2$Co(SeO$_3$)$_2$~\cite{Zhong2020} both realize it, where the latter is closer to the Ising limit~\cite{Chen2026,Zhu2025}. Its low-energy spectrum is dominated by a broad
continuum with roton-like minima at the $M$ point~\cite{Zhu2024,Zhu2025,Chen2026}. Above the gapless excitation at the $K$ point, it
contains a pseudo-Goldstone mode with a gap of about $0.06$\,meV, and no coherent magnon is resolved
within the continuum~\cite{Zhu2025}. Polarized neutron scattering analysis has recently shown
that the transverse ordered moment is small and that the low-energy dynamics is predominantly
longitudinal~\cite{Zhu2026}.

These experimentally measured spectra cannot be described by linear spin-wave theory, which has neither the continuum nor the minimum at
$M$~\cite{Chen2026,Mauri2026,mauri2026-2}. Quantum Monte Carlo
reproduces the measured spectrum of K$_2$Co(SeO$_3$)$_2$, including the $M$ point minima~\cite{Zhu2025}. Tensor network simulations find two low-energy branches: a gapless one, carried by
fluctuations out of the plane of the order, and a gapped pseudo-Goldstone branch, carried by in-plane
fluctuations. Both branches have nearly degenerate minima at $M$~\cite{Gao2024,Chi2024}. These studies attribute the minima to the magnon--Higgs scattering mechanism~\cite{Chi2024,Powalski2015}. Nonlinear
spin-wave theory in the strong easy-axis limit fails to produce the minimum or the small
pseudo-Goldstone gap~\cite{Mauri2026}.
Ref.~\cite{rafael_2025} combines DMRG and several different analytical approaches, demonstrating robust roton feature at the $M$ point.
A roton-like minimum at $M$ is also found in the isotropic triangular antiferromagnet~\cite{Zheng2006PRL,Zheng2006PRB,Ito2017,Macdougal2020}. Its interpretations range from magnon interactions~\cite{Chernyshev2009,Verresen2019} to proximity to a spin liquid with spinon excitations~\cite{Ghioldi2018,Ferrari2019,ZhangLi2020}.

For the easy-axis model, the observed continuum sprectra both numerically and experimentally invite a possible parton description. The natural parent state for a fermionic parton theory is the
$\pi$-flux Dirac spin liquid~\cite{Iqbal2016,Song2019}. Its
Dirac nodes sit at $\pm\bm M/2$, so $\bm M$ is the momentum of inter-valley spinon pairs. Building on
this, Ref.~\cite{Jia2024} described the supersolid of Na$_2$BaCo(PO$_4$)$_2$ as a precursory Dirac spin liquid
at the mean-field level. For the isotropic model with $120^\circ$-order, Ref.~\cite{ZhangLi2020} obtained the roton-like minimum from level repulsion
between the magnon and the spinon continuum, in a fermionic RPA treatment with an on-site Hubbard-type term. There are also other similar fermionic parton constructions in the literature~\cite{ho2001,Jia2024,Bose2025}. The readers can refer to a recent review article on this topic \cite{huang2026review}.

In this work, we use a fermionic parton mean-field
theory built on the $\pi$-flux state, to find the magnetic order self-consistently, and
obtain the collective modes with RPA calculations.
In the parameter regime of interest, the mean-field theory has two self-consistent solutions, $\Y$ and $\Yp$, with $\Yp$ slightly lower in energy. However, including higher order corrections due to fluctuations makes $\Y$ the stable ground state.
On the mean-field level, the $\Y$ state can be stabilized by adding small external fields, after which the excitation spectra can be calculated using RPA. We use two different types of external field: moment-aligned local field and a uniform field along the easy axis. Both cases show stable excitation spectra with a roton minimum at the $M$ point. The former shows a gapped Goldstone mode at the $K$ point due to the explicitly broken $U(1)$ symmetry, whereas the Goldstone of the latter remains gapless because the uniform field preserves the $U(1)$ on the Hamiltonian level. On top of the gapless Goldstone from the transverse channel, there is also a gapped pseudo-Goldstone from the longitudinal channel.
Other than the collective modes at low energy, the parton construction also naturally gives rise to continuum in the excitation spectra.
Since the parton theory proposed here is a phenomenological model, we also investigate the effects of tuning the parameters---the on-site spinon interaction strength and the anisotropy in spin coupling---on the calculated excitation spectra. Comparisons with existing numerical and experimental works are also made.

Note that the parton theory is not \textit{derived} from the original spin model, and should be regarded as a physically motivated phenomenological model with symmetry considerations. 
We remark on the approximations made in this treatment.
First, the single-occupancy constraint is imposed only on average and the trial state is not
Gutzwiller projected. The trial state therefore contains spinon density fluctuations outside of the physical Hilbert space, and the gauge fluctuations of the parton theory are also ignored.
Second, with allowed spinon density fluctuation, a Hubbard-like on-site spinon interaction $U$ is introduced, which penalizes double-occupancy and generates magnetic orders as well as their fluctuations for the trial state. 
As a consequence, the couplings in the model Hamiltonian are phenomenological parameters for the trial state, and should not be directly compared with couplings in the original spin model. Finally, the roton minimum and the continuum in the excitation spectra do not necessarily imply the existence of fractionalized spinons. The purpose here is to provide another possible perspective to the observed physics. Within these limits, the theory gives a self-consistent description of the supersolid order and its low-energy excitation spectrum.

The paper is organized as follows. Sec.~\ref{sec:model} introduces the model, the parton mean-field
theory and the RPA calculation. Sec.~\ref{sec:YYp} compares the two
self-consistent states, $\Y$ and $\Yp$ at the mean-field level. Sec.~\ref{sec:bare} presents the excitation spectra of both states. Sec.~\ref{sec:stab} stabilizes the $\Y$ state with two different types of external field and calculates the excitation spectra of both cases. Sec.~\ref{sec:compare} concludes by comparing our results with recent experiments and numerical studies, and provide some open problems for future study.
\section{Model and parton mean-field theory}\label{sec:model}

\subsection{Spin model}

The Hamiltonian for the spin-$\tfrac12$ XXZ antiferromagnet on the triangular lattice is given by
\begin{equation}
H=\sum_{\langle ij\rangle}\Big[J^z S^z_iS^z_j+J^\perp\big(S^x_iS^x_j+S^y_iS^y_j\big)\Big],
\label{eq:H}
\end{equation}
where in the easy-axis regime $J^z>J^\perp>0$.
In the classical limit the zero-field ground state of Eq.~\eqref{eq:H} is a coplanar
three-sublattice structure whose orientation relative to the easy axis is not fixed by due to energy degeneracy~\cite{Miyashita1986,ShengHenley1992}. Two members of this degenerate family
play the central role below. In the $\Y$ state one sublattice points along the easy axis and the
other two cant symmetrically away from it; in the $\Yp$ state the spin on one sublattice lies in the easy
plane and the other two cant symmetrically away from the lattice plane, which is similar to a $\Y$ state but rotated by $90^\circ$ inside the order plane. Without couplings beyond nearest neighbor, the two have exactly the same classical
energy for every anisotropy~\cite{Gallegos2024}, and the degeneracy is lifted only by quantum
fluctuations, which select $\Y$~\cite{Kleine1992,ShengHenley1992,Gallegos2024} as the true ground state. The $\Y$ state
combines a three-sublattice modulation of $S^z$ that breaks lattice translation (the solid) with in-plane order that breaks the
$U(1)$ spin-rotation symmetry about the $\hat{\bm z}$-axis (the superfluid), and is the spin analogue of the
lattice supersolid of hard-core bosons on the triangular
lattice~\cite{Wessel2005,Heidarian2005,Melko2005}.

\subsection{Parton mean-field theory with the \texorpdfstring{$\pi$}{pi}-flux ansatz}

\begin{figure}[t]
\centering
\includegraphics[width=\columnwidth]{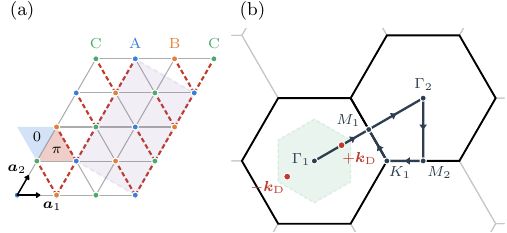}
\caption{Parton construction. (a) Triangular lattice with sublattices $A$, $B$, $C$ and the gauge
used for the $\pi$-flux ansatz: dashed bonds carry hopping of opposite sign, so every up triangle
encloses flux $\pi$ and every down triangle flux $0$. The shaded region is the six-site $\sqrt3\times3$
cell used for the mean-field and RPA calculations; $\bm a_1$, $\bm a_2$ are the primitive vectors.
(b) Brillouin zones of the triangular lattice (solid hexagons, with neighboring zones in gray) and of
the three-sublattice order (shaded), the Dirac nodes $\pm\bm k_{\mathrm D}=\pm\bm M_1/2$ of the spin
liquid, and the path $\Gamma_1$--$M_1$--$\Gamma_2$--$M_2$--$K_1$--$M_1$ along which all spectra in this
paper are shown. With $\bm a_1=a\hat{\bm x}$, the labeled points are $\Gamma_1=0$,
$\bm M_1=(\pi,\pi/\sqrt3)/a$, $\Gamma_2=2\bm M_1$, $\bm M_2=(2\pi,0)/a$ and
$\bm K_1=(4\pi/3,0)/a$.}
\label{fig:model}
\end{figure}

The spin operator on site $i$ can be written in terms of fermionic partons,
$\bm S_i=\tfrac12 f^\dagger_{ia}\bm\sigma_{ab}f_{ib}$, subject to the single-occupancy
constraint $n_i=\sum_a f^\dagger_{ia}f_{ia}=1$, where $a,b$ label the $\up,\dn$ spin states. The parent state is the $\pi$-flux
Dirac spin liquid~\cite{Iqbal2016,Song2019}, in which the nearest-neighbor spinon hopping encloses flux
$\pi$ in every up triangle and $0$ in every down triangle, which can be realized by choosing a spinon hopping phase $\nu_{ij}=\pm1$ across a bond $\langle i,j\rangle$ (see Fig.~\ref{fig:model}(a), where the dashed bonds label $\nu_{ij}=-1$, and 1 otherwise). This staggered flux pattern together with the three-sublattice magnetic order requires the six-site   unit cell shaded in
Fig.~\ref{fig:model}(a), which hosts in total twelve (considering spin degrees of freedom) spinon bands, where the six lowest bands are filled for the mean-field ground state.
The Dirac nodes of the spin liquid sit at $\pm\bm k_{\mathrm D}=\pm\bm M_1/2$
(see Fig.~\ref{fig:model}(b)), giving four (two valley and two spin) two-component Dirac fermions. Magnetic orders will in general gap out the Dirac nodes on the mean-field level, as we discuss below.

We follow Ref.~\cite{ZhangLi2020} and treat the spinon hopping $t$ as an independent
parameter of the trial state, with the magnetic moment $\bm m_i=\langle\bm S_i\rangle$ produced by decoupling the on-site spinon interaction $\sim U\sum_i n_{i\up}n_{i\dn}$ and the nearest-neighbor spin coupling. The
mean-field Hamiltonian is
\begin{equation}
\begin{split}
H_{\mathrm{MF}}={}&-t\sum_{\langle ij\rangle,a}\nu_{ij}\big(f^\dagger_{ia}f_{ja}+\mathrm{h.c.}\big)\\
&+\sum_{i,a,b}\bm h_i\cdot\left(\tfrac12 f^\dagger_{ia}\bm\sigma_{ab}f_{ib}\right)+\sum_i\lambda_i\,(n_i-1),
\end{split}
\label{eq:HMF}
\end{equation}
with the local field
\begin{equation}
\bm h_i=\bm h^{\mathrm{ex}}_i-U\bm m_i,\quad
h^{\mathrm{ex},\alpha}_i=J^\alpha\sum_{j\in\partial i}m^\alpha_j,
\label{eq:field}
\end{equation}
where $J^x=J^y=J^\perp$, $\partial i$ denotes the six neighbors of site $i$, and $\alpha =x,y,z$. The Lagrange multipliers $\lambda_i$ implement the on-site unit filling on average for the spinons. Notice that to avoid clutter of notations, here we also use $J^{x,y,\perp}$ for the couplings as in the original spin model in Eq.~(\ref{eq:H}), but we emphasize that their actual numerical values can be quite different due to the phenomenological nature of the parton mean-field theory in Eq.~(\ref{eq:HMF}). The exchange mean field
$\bm h^{\mathrm{ex}}_i$ comes from the nearest-neighbor coupling, and the term $U\bm m_i$ comes from decoupling the spinon interaction term in the spin channel.
The mean-field trial state can be tuned across the transition from the spin liquid to the ordered phase through these effective Weiss fields. As we will show below, in our ``mean-field + RPA'' framework, the spinon interaction $U$ is essential to give rise to a sharp roton signature in the excitation spectrum. As a comparison, Appendix~\ref{app:noU} shows the construction without the onsite $U$ term, where the orders come from the exchange coupling alone. Also note that the anisotropy enters the trial state only through $\bm h^{\mathrm{ex}}_i$ while the hopping and the $U$ term are isotropic, so the anisotropy ratio $J^\perp/J^z$ of Eq.~\eqref{eq:H} need not coincide with that of the original spin model.

We diagonalize $H_{\mathrm{MF}}$, with the mean-field ground state given by filling the lowest six bands, and solve the on-site magnetic orders self-consistently.
The solutions are stationary points of the following energy functional
\begin{equation}
E_{\mathrm{MF}}[\bm m]=\frac1N\Big[\sum_{n\bm k\in\mathrm{occ}}\varepsilon_{n\bm k}(\bm m)
-\frac12\sum_i\bm h_i\cdot\bm m_i\Big],
\label{eq:EMF}
\end{equation}
where $\varepsilon_{n\bm k}$ are the eigenvalues of $H_{\mathrm{MF}}$, and the second term is due to double-counting in mean-field decoupling.

\subsection{Collective modes}

\begin{figure}[t]
\centering
\includegraphics{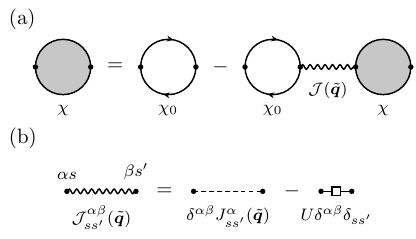}
\caption{Diagrams of the RPA. (a) Dyson equation for the RPA susceptibility $\chi$ (shaded bubble),
Eq.~\eqref{eq:rpa}. An open bubble is the bare susceptibility $\chi_0$: its solid lines are the
propagators of the mean-field spinons of Eq.~\eqref{eq:HMF} in the ordered state, and its dots are
the spin vertices $\tfrac12\sigma^\alpha$ on site $s$ of the six-site cell. The sign convention
$\delta\langle\bm S\rangle=-\chi_0\,\delta\bm h$ absorbs the sign of the fermion loop into $\chi_0$.
(b) The kernel $\Jt(\tilde{\bm q})$ of Eq.~\eqref{eq:kernel} (wavy line) is the exchange
$J(\tilde{\bm q})$ (dashed), minus the on-site vertex $U$
(square).}
\label{fig:rpadiag}
\end{figure}

The fluctuations around the mean-field state, treated here using RPA, give rise to collective modes, which are essentially coherent spinon particle-hole excitations. Let $\chi_0(\tilde{\bm q},\omega)$
be the $18\times18$ bare spin susceptibility of the mean-field spinons, with
indices running over the three spin components (labeled by $\alpha,\beta$) and the six sites of the unit cell (labeled by $s,s'$), and with the sign
chosen so that $\delta\langle\bm S\rangle=-\chi_0\,\delta\bm h$. Here the tilde in $\tilde{\bm q}$ is to denote the momentum in the smaller folded Brillouin zone (BZ) due to the enlarged real-space unit cells from magnetic order and flux pattern. The momentum $\bm q$ in the original BZ of the triangular lattice can be identified with a $\tilde{\bm q}$ up to a reciprocal lattice vector $\tilde{\bm G}$ of the folded BZ.
The bare $\chi_0$ is given by
\begin{equation}
    \chi_{0,Is,I's'}^{\alpha\beta}(\tau)=\la \mathcal{T}\delta S^\alpha_{Is}(\tau)\delta S^\beta_{I's'}(0) \ra,
\end{equation}
where the indices $I,I'$ label the enlarged unit cells, $\tau$ is the imaginary time with $\mathcal{T}$ denoting time-ordering and $\delta S^\alpha\equiv S^\alpha-\la S^\alpha \ra$. Fourier transform in time and in space with respect to $I$ gives
\begin{equation}
    \chi_{0,ss'}^{\alpha\beta}(\tilde{\bm q},i\omega)=\sum_I\int d\tau \e^{i\omega\tau-i\bm R_I\cdot\tilde{\bm q}}\chi_{0,Is,0s'}^{\alpha\beta}(\tau),
\end{equation}
where $\bm R_I$ is the coordinate for the cell $I$, and the cell $I'$ is translated to the origin. In the mean-field parton theory, the bare $\chi_0$ is a single fermion particle-hole bubble diagram shown in Fig.~\ref{fig:rpadiag}(a).
The RPA susceptibility is
\begin{equation}
\chi(\tilde{\bm q},i\omega)=\big[1+\chi_0(\tilde{\bm q},i\omega)\Jt(\tilde{\bm q})\big]^{-1}\chi_0(\tilde{\bm q},i\omega),
\label{eq:rpa}
\end{equation}
with the kernel
\begin{equation}
\Jt(\tilde{\bm q})=J(\tilde{\bm q})-U\,\openone ,
\label{eq:kernel}
\end{equation}
where $J(\tilde{\bm q})$ is the Fourier transform of the anisotropic exchange matrix on the six-site cell.
Diagrammatically, Eq.~\eqref{eq:rpa} sums the chain of spinon particle--hole bubbles joined by the
kernel (Fig.~\ref{fig:rpadiag}(a)), which has two parts: the exchange coupling that connects neighboring
sites, and the on-site interaction vertex (Fig.~\ref{fig:rpadiag}(b)). The susceptibility at wave vector
$\bm q$ is obtained by unfolding with the sublattice positions $\bm r_s$,
\begin{equation}
\chi^{\alpha\beta}(\bm q,i\omega)=\frac16\sum_{s,s'}e^{-i\bm q\cdot(\bm r_s-\bm r_{s'})}
\chi^{\alpha\beta}_{ss'}(\tilde{\bm q},i\omega),
\end{equation}
after which the measurable dynamical spin structure factor, characterizing retarded physical response, is readily obtained by analytic continuation \begin{equation}
S^{\alpha\beta}(\bm q,\omega)=\frac{1}{\pi}\mathrm{Im}\,\chi^{\alpha\beta}(\bm q,i\omega\to \omega+i0^+).
\end{equation}

Local stability of the ground state can be read off the eigenvalue signs of the static RPA corrected susceptibility $\chi(\tilde{\bm q},0)$, which are further related to the eigenvalue signs of the denominator $1+\chi_0(\tilde{\bm q},0)\Jt(\tilde{\bm q})$ in Eq.~(\ref{eq:rpa}), since $\chi_0$ has non-negative eigenvalues. Therefore, we can just focus on the denominator. A negative eigenvalue marks a direction in which the
mean-field energy can further decrease under fluctuations, indicating an instability.

\section{Self-consistent solutions}\label{sec:YYp}
\begin{figure}[t]
\centering
\includegraphics[width=\columnwidth]{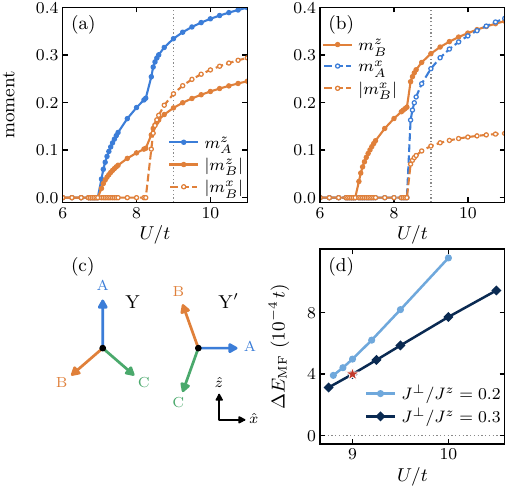}
\caption{Onset of order and energetics of the $\Y$ and $\Yp$ states at $J^z/t=2$. (a) Sublattice
moments of the $\Y$ solution against $U/t$ at $J^\perp/J^z=0.3$: $m^z_A$, $|m^z_B|$ and the
transverse moment $|m^x_B|$; the $C$ moment mirrors $B$. The dotted line marks $U/t=9$, and is used for later calculations. (b) Sublattice moments of the $\Yp$ solution. (c) The two self-consistent solutions at $J^\perp/J^z=0.3$, $U/t=9$ (star in
(d)), with the moments of the three sublattices drawn from a common origin, to scale, in the $xz$ plane,
with $\hat{\bm z}$ the easy axis. Colors label the sublattices as in Fig.~\ref{fig:model}. (d) Mean-field splitting
$\Delta E_{\mathrm{MF}}=E_{\mathrm{MF}}(\Y)-E_{\mathrm{MF}}(\Yp)$ per site for $J^\perp/J^z=0.2$ and $0.3$. Since $\Delta E_\mathrm{MF}>0$, $\Yp$ is lower in energy.}
\label{fig:energy}
\end{figure}

\begin{figure*}[t]
\centering
\includegraphics[width=\textwidth]{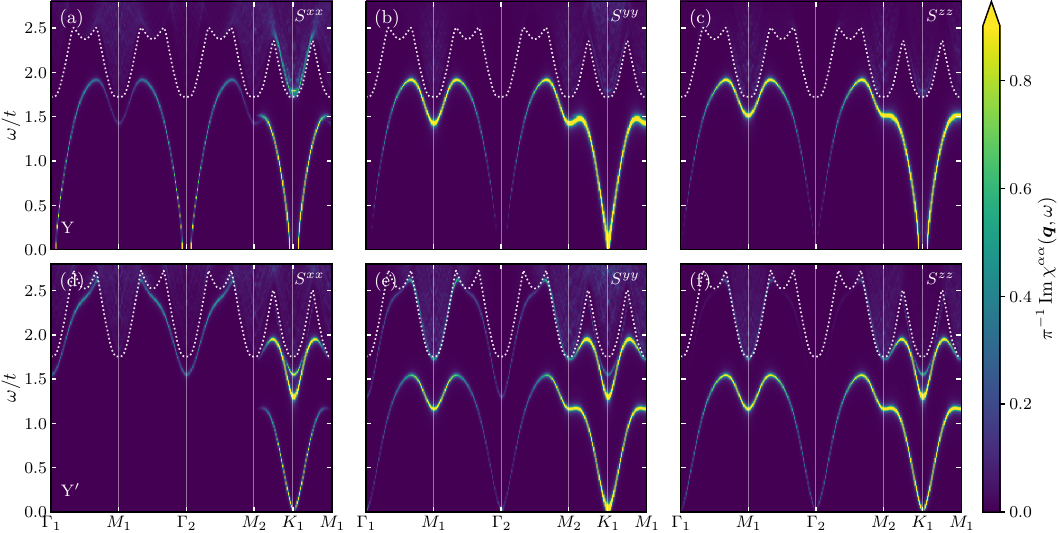}
\caption{Dynamical structure factor $S^{\alpha\alpha}(\bm q,\omega)=\pi^{-1}\mathrm{Im}\,\chi^{\alpha\alpha}(\bm q,\omega+i0^+)$
of the two self-consistent states, along the path
$\Gamma_1$--$M_1$--$\Gamma_2$--$M_2$--$K_1$--$M_1$ of Fig.~\ref{fig:model}(b), at the parameters $J^z/t=2$, $J^\perp/J^z=0.3$, $U/t=9$. (a)-(c) The unstable $\Y$ state. (d)-(f) The stable $\Yp$ state. Columns show
$\alpha=x,y,z$; both states are ordered in the $xz$ plane. The dotted line is the lower edge of the
two-spinon continuum. The spectra use a Lorentzian broadening $\eta=0.015\,t$ and $\chi_0$ on a
$16\times16$ momentum space mesh of the six-site cell; the continuum edge uses a $24\times24$ mesh.}
\label{fig:bare}
\end{figure*}

\subsection{Two self-consistent states: \texorpdfstring{$\Y$ versus $\Yp$}{Y versus Y'}}

At fixed anisotropy, increasing $U$ first produces a collinear solid, in which only the $z$
components order, and transverse order sets in only at a somewhat larger $U$ (see Fig.~\ref{fig:energy}(a)). At $J^z/t=2$ and $J^\perp/J^z=0.3$ the transverse order appears just below $U/t=8.5$. Our
working parameter set is $J^z/t=2$, $J^\perp/J^z=0.3$, $U/t=9$. For comparison we also use
$J^\perp/J^z=0.2$ at the same $J^z/t$ and $U/t$. More detailed analysis on the effects of tuning the $U$ and the anisotropy are provided in Appendix~\ref{app:uj} and \ref{app:aniso}.

Two self-consistent solutions coexist for the chosen parameters, shown in
Fig.~\ref{fig:energy}(b). The first is the $\Y$ state: $\bm m_A=(0,0,0.334)$ along the easy axis and
$\bm m_{B,C}=(\mp0.218,0,-0.189)$, of length 0.288, so that $\bm m_B$ and $\bm m_C$ make an angle of
$98.3^\circ$ with each other and $130.8^\circ$ with $\bm m_A$. The moments are reduced well below the
classical value $1/2$ and leave a small net magnetization
$\bm M_F=1/3\sum_s\bm m_s=(0,0,-0.014)$. The second self-consistent solution is the $\Yp$ state, in
which the $\bm m_A$, of length 0.271, lies in the lattice plane and the other two, of length 0.321,
sit $19.7^\circ$ from $+\hat{\bm z}$ and $-\hat{\bm z}$. Its net moment, $\bm M_F=(0.018,0,0)$, points in the
easy plane. Both states are coplanar in the $xz$ plane and break the $U(1)$ symmetry and the lattice
translations. They are the parton mean-field solutions to the quantum counterparts of the classical $\Y$ and $\Yp$ states \cite{Gallegos2024,Jiang2023}.

\subsection{Mean-field energetics}

The mean-field energy favors $\Yp$, shown in Fig.~\ref{fig:energy}(d). 
At the chosen parameters, the calculated mean-field energy density $\Delta E_{\mathrm{MF}}\equiv E_{\mathrm{MF}}(\Y)-E_{\mathrm{MF}}(\Yp)>0$. 
As Fig.~\ref{fig:energy}(d) shows, the
preference for $\Yp$ is not specific to one parameter set: it holds throughout a finite region in parameter space.
The energy difference between the two states increases further with the growth with $U$ can be understood as the following. Among all the couplings, $U$ is the dominant one. Given
\begin{equation}
\frac{\pd E_{\mathrm{MF}}}{\pd U}=-\frac1{2N}\sum_i|\bm m_i|^2 ,
\label{eq:HF}
\end{equation}
the energy difference mainly comes from the sum of the squared moments of the two states. The $\Yp$ state
has $N^{-1}\sum_i|\bm m_i|^2=0.0933$, in comparison to $0.0926$ for $\Y$, leading to a lower energy for $\Yp$.  
The fact that the mean-field theory fails to pick out the true ground state parallels a similar problem in the classical model, where the two states are degenerate in terms of classical energy, and the selection has to come
from fluctuations~\cite{Kleine1992}. We explicitly consider the effects of fluctuations, namely the zero-point energy, in Appendix~\ref{app:zero-point}. It is shown there that the $\Y$ state can become lower in energy once the zero-point energy contribution is included.

\section{Dynamical spin susceptibility}\label{sec:bare}

\begin{figure}[t]
\centering
\includegraphics[width=\columnwidth]{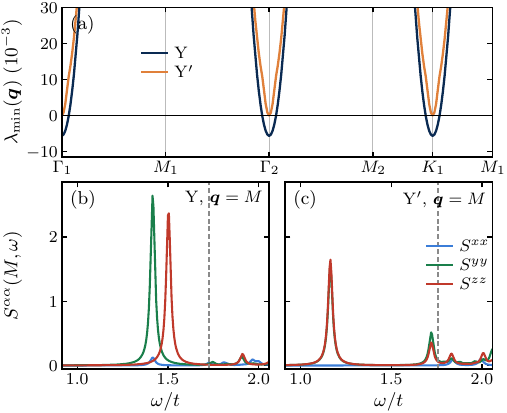}
\caption{Stability and the structure factor at the $M$ point for both states at $J^z/t=2$, $J^\perp/J^z=0.3$,
$U/t=9$. (a) Smallest eigenvalue $\lambda_{\min}(\bm q)$ of $1+\chi_0(\bm q,0)\Jt(\bm q)$
along the path of Fig.~\ref{fig:bare}, for $\Y$ and $\Yp$. (b) $S^{\alpha\alpha}(M,\omega)$ of $\Y$ in
the three channels. (c) The same for $\Yp$. Dashed lines mark the continuum edge at $M$. All panels use
a Lorentzian broadening $\eta=0.015\,t$ and a $16\times16$ momentum space mesh for $\chi_0$, and a $24\times24$ mesh for the continuum edge calculation.}
\label{fig:baremodes}
\end{figure}

Here we compute the RPA corrected spin susceptibility of both self-consistent states found above. Fig.~\ref{fig:bare} shows $S^{\alpha\alpha}(\bm q,\omega)$ along the
path $\Gamma_1$--$M_1$--$\Gamma_2$--$M_2$--$K_1$--$M_1$ of Fig.~\ref{fig:model}(b) at the chosen parameters, for $\Y$ in the top row
and $\Yp$ in the bottom one. Fig.~\ref{fig:baremodes}(a) shows, along the same path, the smallest
eigenvalue $\lambda_{\min}(\bm q)$ of the static matrix $1+\chi_0(\bm q,0)\Jt(\bm q)$. As mentioned before, for a stable state all the eigenvalues should be non-negative. Therefore, a negative $\lambda_\mathrm{min}$ indicates instability.

\subsection{Stability and excitation spectra}

The $\Yp$ state is stable, its $\lambda_{\min}(\bm q)$ is positive at all $\bm q$ except for vanishing at the $\Gamma$ and $K$ points. These are the zone center of the six-site cell, where the Goldstone mode, and the corresponding gapless cone is visible at $K_1$ in
Figs.~\ref{fig:bare}(d)-(f). For $\Y$, in contrast, $\lambda_{\min}(\bm q)$ turns negative inside small regions around the zone center.
The corresponding excitation spectra $S^{xx}$ and $S^{zz}$ in Figs.~\ref{fig:bare}(a)(c) show the same instability.
Nonetheless, both show a pronounced minimum of the
collective mode at $M$ point. In Sec.~\ref{sec:stab}, we use external fields to stabilize the $\Y$ state and recalculate the spectra.

\subsection{The roton minimum at \texorpdfstring{$M$}{M}}\label{sec:whyM}

The roton minimum at the $M$ point can be understood in the spinon picture as the following. The Dirac nodes of the parent
spin liquid sit at $\pm\bm M/2$, so an excited spinon particle-hole pair across the two valleys carries
momentum around $\bm M$ and energy about twice the spin gap. The lower edge of the continuum therefore dips
at the $M$ points (dotted lines in Fig.~\ref{fig:bare}). Due to level-repulsion, the collective modes, when exist, are pushed
down most strongly at $M$. 
In our model, the phenomenological parameter $U$ has to be sufficiently large to set the continuum to a high energy such that the collective modes at lower energy are not strongly damped and remain sharp (see Appendix~\ref{app:noU} and Appendix~\ref{app:uj} for more details). On the other hand, $U$ cannot be too strong, otherwise the continuum would be too high up in energy to repel the low-energy collective mode at $M$ into a local minimum~\cite{ZhangLi2020}. Figs.~\ref{fig:baremodes}(b)(c) further resolve the mode amplitudes at $M$ for different channels. For the $\Y$ state, there are two modes at $M$, where the transverse $S^{yy}$ component has the lowest energy and the longitudinal $S^{zz}$ has slightly higher energy. For $\Yp$, the two modes are nearly degenerate. The $S^{xx}$ contribution to both states at $M$ is insignificant.

\section{Stabilizing the Y state}\label{sec:stab}
\begin{figure}[t]
\centering
\includegraphics[width=\columnwidth]{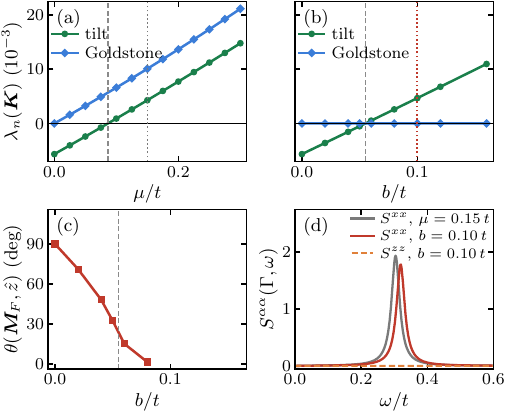}
\caption{Stabilization of the $\Y$ state at $J^z/t=2$, $J^\perp/J^z=0.3$, $U/t=9$.
(a) The two smallest eigenvalues $\lambda_n(\bm K)$ of $1+\chi_0(\bm K,0)\Jt(\bm K)$ as a function of the
moment-aligned field $\mu$. The lower branch is the tilt mode and
the upper one the Goldstone mode, which $\mu$ gaps when $\mu\neq 0$ due to explicit symmetry breaking. The dashed line labels $\mu_c\approx 0.086\,t$ and the dotted line marks $\mu=0.15\,t$, used in
Figs.~\ref{fig:spectra}(a)--(c). (b) The two smallest eigenvalues against the uniform field . The dashed line marks and the dotted line
$b=0.10\,t$, used in Figs.~\ref{fig:spectra}(d)--(f). (c) Angle between the net moment $\bm M_F$ and $\hat{\bm z}$ for the self-consistency solution with an initial $\Yp$ state; it rotates into a $\Y$ state with the increase of $b$.
 (d) Low-energy $S^{xx}(\Gamma,\omega)$
for $\mu=0.15\,t$ and $b=0.10\,t$, and $S^{zz}(\Gamma,\omega)$ for $b=0.10\,t$, which vanishes because
$S^z_{\mathrm{tot}}$ is conserved; $\eta=0.015\,t$ and a $16\times16$ momentum space mesh.}
\label{fig:stab}
\end{figure}

The inclusion of the zero-point energy makes the $\Y$ state the stable ground state, but at the mean-field level $\Yp$ is the stable one. In the following we apply a small external field to stabilize the $\Y$ state to compute its excitation spectrum. We do it in two different ways, and in both approaches the applied field is external in the sense that it is independent of the magnetic order, so they leave the $\Jt$ of Eq.~\eqref{eq:kernel} unchanged.

\begin{figure*}[t]
\centering
\includegraphics[width=\textwidth]{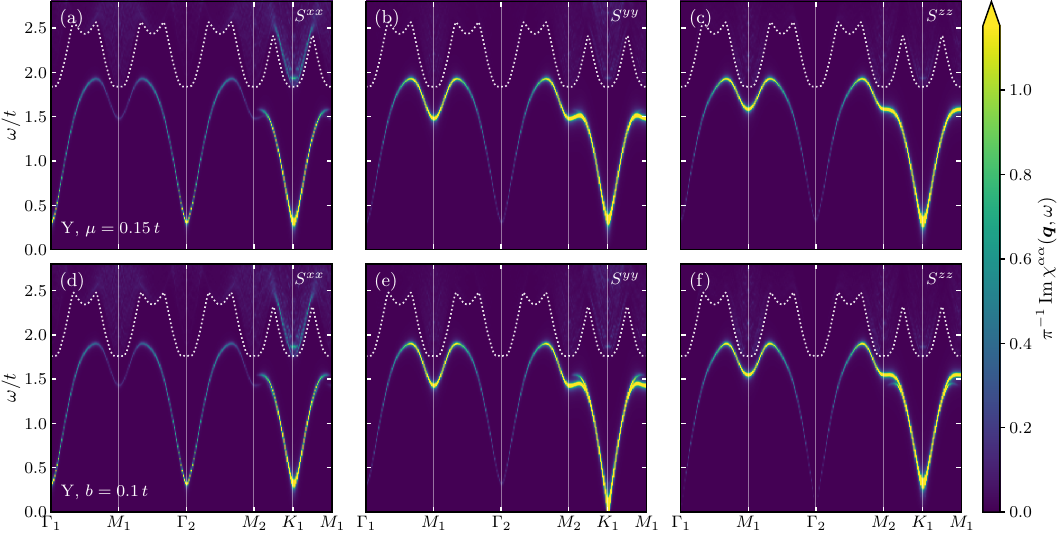}
\caption{Dynamical structure factor $S^{\alpha\alpha}(\bm q,\omega)=\pi^{-1}\mathrm{Im}\,\chi^{\alpha\alpha}(\bm q,\omega)$
of the stabilized $\Y$ state along $\Gamma_1$--$M_1$--$\Gamma_2$--$M_2$--$K_1$--$M_1$, at
$J^z/t=2$, $J^\perp/J^z=0.3$, $U/t=9$. (a)-(c) With the moment-aligned field
$\mu=0.15\,t$ of Eq.~\eqref{eq:mu}. (d)-(f) With the uniform field $b=0.10\,t$ of
Eq.~\eqref{eq:b}.}
\label{fig:spectra}
\end{figure*}

\subsection{Moment-aligned field}\label{sec:mu}

The first approach follows the modified spin-wave theory of Ref.~\cite{Coletta2013}, used to tackle the $\Y$--$\Yp$ competition in Ref.~\cite{Gallegos2024} and adds
\begin{equation}
\delta H_{\mu}=\mu\sum_i\left(\lvert\bm m_i\rvert-\hat{\bm m}_i\cdot\bm S_i\right),
\label{eq:mu}
\end{equation}
which acts as a field of strength $\mu$ along the local ordered direction $\hat{\bm m}_i$ on each site. We keep the moments
$\bm m_i$ of the $\mu=0$ solution and re-solve only the multipliers $\lambda_i$, so that the filling
stays on average at one spinon per site. $\langle\delta H_\mu\rangle$ vanishes in
the reference state, so the energy can only change at higher orders in $\mu$. Since the ordered moments are kept fixed, the term $-\hat{\bm m}_i\cdot \bm S_i$ acts as a restoring potential for the moments, stabilizing the $\Y$ state. Another way of looking at this is that introducing the $\delta H_\mu$ makes the $\Y$ state more rigid to spin flips, which is reflected in a decreasing magnitude of the bare susceptibility $\chi_0$. As a result, the eigenvalues of $1+\chi_0\Jt$ can be shifted to non-negative (stable) values. 

Fig.~\ref{fig:stab}(a) shows the two smallest eigenvalues at $\bm K$ as $\mu$ increases.
The one that starts from 0 at $\mu=0$ corresponds to the gapless Goldstone mode, whereas the initially negative one at $\mu=0$ represents the unstable tilt mode that rotates the $\Y$ state towards the stable $\Yp$ carrying lower energy. Both grow linearly with $\mu$. Because $\delta H_\mu$ breaks the $U(1)$ symmetry
explicitly, the Goldstone mode acquires a gap. We select $\mu=0.15\,t$ (dotted line), at which no eigenvalue is negative at any wave vector along the path shown in Fig.~\ref{fig:spectra}.
The resulting spectra are shown in Figs.~\ref{fig:spectra}(a)-(c).
\subsection{Uniform field along the easy axis}
\label{sec:b}
The second option is a small uniform field along the easy axis,
\begin{equation}
\delta H_b=b\sum_iS^z_i ,
\label{eq:b}
\end{equation}
with $b>0$, so that the field favors moments along $-\hat{\bm z}$: it is antiparallel to $\bm m_A$ and
parallel to the net moment of $\Y$. Unlike $\mu$ in Eq.~(\ref{eq:mu}), this field preserves
the $U(1)$ symmetry of Eq.~\eqref{eq:H} and the Goldstone mode stays exact (Fig.~\ref{fig:stab}(b)). It also directly contributes to the energy, which leads to a self-consistent $\Y$ state with a lower energy compared to the $\Yp$, and the RPA response corresponds to the physical response of the model in a longitudinal field.

The stabilization can be understood as the following. At first order in $b$ the field lowers the energy of $\Y$, whose net moment lies
along $\hat{\bm z}$, in contrast to $\Yp$, whose net moment is perpendicular to it.
As shown in Fig.~\ref{fig:stab}(c), under the increasing external field the net moment of $\Yp$ rotates
continuously from the lattice plane toward $\hat{\bm z}$, and eventually converges to the $\Y$ state.
We choose $b=0.1\,t$ (dotted line in Fig.~\ref{fig:stab}(b)), and the excitation spectra are
shown in Figs.~\ref{fig:spectra}(d)-(f). The Goldstone cone remains gapless at $K$, and $S^{zz}$
vanishes at $\Gamma$ as required by the conservation of $S^z_{\mathrm{tot}}$. The field also increases
the net moment from $0.014$ to $0.018$, but its effect on the structure of the $\Y$ order and the modes at $M$ is mild.

For both approaches used to stabilize the $\Y$ state, the roton feature at $M$ is robust. Another noticeable feature that can be seen in both Fig.~\ref{fig:stab}(d) and the excitation spectra in Fig.~\ref{fig:spectra} is the appearance of a bright mode in the $S^{xx}$ component at $\Gamma_2$. This is the gapped \textit{global} rotation mode away from the stabilized $\Y$ state, which rotates the net moment along the $\hat{z}$ direction towards the $\hat{\bm x}$ direction around the $y$-axis inside the order plane. During such infinitesimal rotations, the $y$-component of the order remains the same, and the $z$-component changes only by second order with the rotation angle, in contrast to the first order change for the $x$-component. Therefore, the tilt mode does not show up in $S^{yy}$ or $S^{zz}$.

\section{Discussion}\label{sec:compare}
We have studied the supersolid $\Y$ state of the spin-$\tfrac12$ easy-axis XXZ antiferromagnet on the triangular lattice in a
fermionic parton mean-field theory built on the $\pi$-flux ansatz, augmented by the RPA calculations. The phenomenological model presented here is able to capture some of the main low-energy spectroscopic features of the spin supersolid observed experimentally.
Below we provide a comprehensive comparison between this work and the existing experimental and numerical studies. Due to the phenomenological nature of our theoretical model, we compare the observable response features rather than absolute
energy scales. We summarize both points of agreement and points of disagreement, and then discuss some open directions.

The most robust common feature is the minimum at $M$. All neutron studies of
K$_2$Co(SeO$_3$)$_2$ find a roton-like minimum there~\cite{Chen2026,Zhu2026}. Quantum Monte Carlo
reproduces the measured spectrum, minima included~\cite{Zhu2025}, and DMRG and
tensor networks find the minimum at $ M$ as well~\cite{Gao2024,Chi2024,rafael_2025}. 
Importantly, this feature goes beyond the description of spin-wave theory:
linear spin-wave theory of the nearest-neighbor model has no minimum at $M$,
and in the strong easy-axis limit one-loop corrections cannot produce one together with a small
pseudo-Goldstone gap~\cite{Mauri2026}.

The structure of the low-energy modes also agrees well between our calculation and other heavy numerical treatments. 
DMRG and tensor networks find two low-energy branches: a
gapless one, carried by fluctuations out of the plane of the order and containing the Goldstone mode,
and a gapped pseudo-Goldstone branch carried by in-plane fluctuations. Both have a minimum at $M$,
where they are nearly degenerate~\cite{Gao2024,Chi2024}.
Neutron scattering resolves a gapped mode at $0.06$\,meV next to the gapless excitations, and in small
fields finds the two branches nearly degenerate at $M$~\cite{Chen2026,Zhu2025}. The parton theory also has two branches. In the field-stabilized $\Y$
state (Figs.~\ref{fig:spectra}(d)-(f)), the $S^{yy}$ branch rises from the Goldstone mode at $K$
and comes down to a local minimum at $M$, while the branch carried by $S^{zz}$ starts from
the gap from the rotation mode at $K$ and ends at slightly higher energy at $M$. As in the
tensor-network calculation, the lower mode at $M$ is the out-of-plane one.

The response to a longitudinal field of our model agrees qualitatively with experiments. In K$_2$Co(SeO$_3$)$_2$ the pseudo-Goldstone
gap at $K$ grows from $0.06$\,meV in zero field to $0.16$\,meV and $0.26$\,meV at $0.25$\,T and $0.5$\,T, respectively. The
minimum at $M$ in the gapless branch is progressively suppressed, while the dip in the gapped branch
survives at $0.25$\,T~\cite{Zhu2025}. These are predicted by tensor-network
calculation, attributing the minimum of the gapless branch to level repulsion from the gapped
one~\cite{Chi2024}. In our calculation, applying the magnetic field along the $z$-direction also tends to flatten out the roton minimum at $M$.
Two further points agree. The zero-point energy selects $\Y$ over $\Yp$, as in spin-wave
theory and DMRG~\cite{Kleine1992,Gallegos2024}. The broad continuum in the measured
spectrum has a natural counterpart in the two-spinon continuum of the parton theory.

Regarding the points of disagreement, one point concerns the polarization of the low-energy weight. Ref.~\cite{Zhu2026} finds the lowest-energy branch of K$_2$Co(SeO$_3$)$_2$ to be
longitudinal, whereas in our case it is transverse. The authors attribute the
longitudinal character to the proximity of the Ising limit. There the low-energy excitations connect
states within the degenerate manifold of Ising ground states and conserve $S^z$, while transverse
excitations cost an energy of order $J^z$~\cite{Zhu2026}. This mechanism has no counterpart in our trial state, whose transverse order is generated mainly by an effective isotropic Weiss mean-field.
The ordered moments differ as well. Polarized neutron scattering gives
$\langle S^\perp\rangle/\langle S^z\rangle\approx0.32$~\cite{Zhu2026}, larger than
the DMRG value of about $0.05$ at $J^\perp/J^z=0.07$~\cite{Gallegos2024}. Exact diagonalization and
tensor networks, by contrast, suggest that the transverse moment may vanish close to the Ising limit~\cite{Ulaga2024,Ulaga2025,Xu2025}.
The corresponding ratio for our chosen parameters is about twice the measured one. Our $\Y$ state also carries a net moment of about $2.9\%$ of saturation, whereas the quantum
$\Y$ state is expected to carry none~\cite{Zhu2026}.

The aforementioned disagreements point to the next steps. One common issue with the parton construction discussed here is that the on-site half-filling is only implemented on average.
Therefore, the Gutzwiller projection, enforcing on-site half-filling constraint, through the variational Monte Carlo algorithm built on the same $\pi$-flux ansatz is a natural direction~\cite{Ferrari2019}.
Another interesting problem that can be studied is the effect of a next-nearest-neighbor $J_2$ coupling in the spin model~\footnote{We thank Johannes Knolle for bringing up this point}, which plays a crucial role in stabilizing the Dirac spin liquid~\cite{Gallegos2024,willsher2025dynamics}, even though the formation of the supersolid phase does not seem to require the $J_2$. Furthermore, a potential issue is the usage of the $\pi$-flux Dirac spin liquid as the starting point. As shown in Ref.~\cite{Gallegos2024}, the spin liquid phase occupies a region in the phase diagram with $J_2\neq 0$. Therefore, the $J_2=0$ case is not directly in proximity to the spin liquid. This calls into the question whether the $\pi$-flux ansatz is sufficiently justified, even though the particle-hole excitation across the two Dirac nodes naturally gives rise to excitations carrying momentum $\bm M$ (see Sec.~\ref{sec:whyM}). 
This aspect of the parton theory also deserves further investigation.

\begin{acknowledgments}
We thank Y-Z. You, L. Su, R. Flores-Calderón, M. Ulaga, J. Knolle, J. Willsher, S. A. Chen, C. D. Batista, J. Walkling, U. Seifert, H. Yan, H. Zhao, Y. Hou, N. Chakraborty, D. Chowdhury, T. Chen, L. M. Chinellato for discussions. This work is supported in part by the Alexander von Humboldt Foundation through a Humboldt Research Fellowship, and the Leibniz Association through Leibniz Competition Project No. J200/2024.
\end{acknowledgments}
\appendix
\section{Zero-point energy}\label{app:zero-point}
\begin{figure}[t]
\centering
\includegraphics{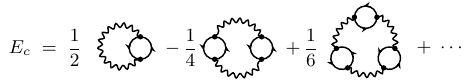}
\caption{Ring-diagram expansion of the RPA zero-point energy, Eq.~\eqref{eq:Ec}, where the coefficients contain exactly the symmetry factors of the diagrams.}
\label{fig:ringdiag}
\end{figure}
\begin{figure}
    \centering
    \includegraphics[width=0.95\linewidth]{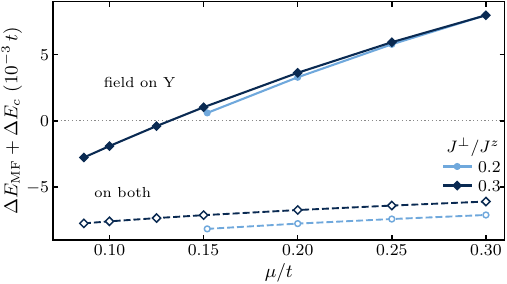}
    \caption{Energy difference between $\Y$ and $\Yp$ when the zero-point energy is taken into consideration. Solid markers show the results for the case where the moment-aligned field $\mu$ is only applied to $\Y$ for stabilization, and the open markers show the results for the case where the field is applied to both states as a comparison. Results for two different anisotropies are shown, where a bigger $\mu$ is required to stabilize the $\Y$ for $J^\perp/J^z=0.2$.}
\label{fig:zero-point}
\end{figure}
The quantity that the mean field leaves out is the zero-point energy of the collective modes, the
parton analogue of the $\frac12\sum_{\bm q}\omega_{\bm q}$ term of spin-wave theory, where $\omega_{\bm q}$ is the spin-wave dispersion~\cite{Gallegos2024,ShengHenley1992}. At the level of
the RPA it is the ring sum
\begin{equation}
\begin{split}
E_c=\frac{1}{2N}\sum_{\tilde{\bm q}}\int_{-\infty}^\infty\!\frac{d\omega}{2\pi}\;
\mathrm{Tr}\Big[\ln\big(1+\chi_0(\tilde{\bm q},i\omega)\Jt(\tilde{\bm q})\big)\Big]
\end{split}
\label{eq:Ec}
\end{equation}
Expanding the
logarithm turns Eq.~\eqref{eq:Ec} into the sum of the ring diagrams of Fig.~\ref{fig:ringdiag}, where each boson line carries the same
$(\tilde{\bm q},i\omega)$, and the diagram with $n$ fermion loop comes with a symmetry factor of $(-1)^{n+1}/2n$.

The calculation of the zero-point energy is only well-defined when the mean-field state is stable. Therefore, we use the small moment-aligned field discussed in the main text to stabilize the $\Y$ state in the $E_c$ calculation. The stable $\Yp$ state needs no such stabilization.
The results are shown in Fig.~\ref{fig:zero-point}. For sufficiently small $\mu$, including the fluctuation contributions reverses the mean-field energy ordering, making $\Y$ the lower energy state. The logic is similar to that in the modified spin-wave treatment of Ref.~\cite{Gallegos2024}, where $\Yp$ is
lower classically for antiferromagnetic next-nearest-neighbor coupling $J_2$ and $\Y$ is recovered once the zero-point energy is
added. 
As a comparison, the case where a $\mu$ field is added to both states is also presented in Fig.~\ref{fig:zero-point} shown in open markers. In this case, the $\Y$ state has lower energy for the whole range of $\mu$ investigated. 

\section{Without on-site spinon interaction}\label{app:noU}
\begin{figure}[t]
\centering
\includegraphics[width=\columnwidth]{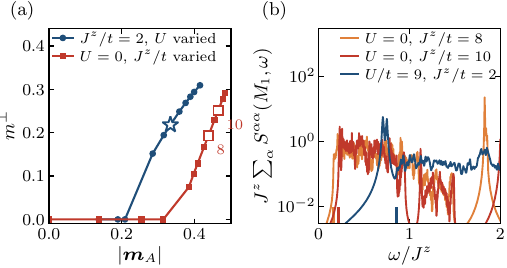}
\caption{The construction with and without the $U$, at $J^\perp/J^z=0.3$.
(a) Transverse moment $m^\perp$ of the $\Y$-seeded solution against its Ising moment $|\bm m_A|$,
with $U$ ($J^z/t=2$, $U$ increasing) and without $U$ ($U=0$,
$J^z/t$ increasing, from $4.5$ to $14$). The star marks the $U=9$ point used in the main text, the open squares the points
$J^z/t=8$ and $10$ of Fig.~\ref{fig:noUspec}. (b) $\sum_\alpha S^{\alpha\alpha}(\bm M_1,\omega)$ in units
of $1/J^z$, against $\omega/J^z$ (for better comparison, since the scale of the collective modes are set by the exchange coupling). Short vertical bars mark the continuum edge at $M$.}
\label{fig:noU}
\end{figure}

In this section we set $U=0$ in Eqs.~\eqref{eq:field} and \eqref{eq:kernel}, so that the exchange
field alone has to produce the moment, and keep $J^\perp/J^z=0.3$. The only remaining
parameter is $J^z/t$. The numerical settings are those of the main text, except where stated.

\subsection{Self-consistency solutions}

Without $U$ the spin liquid is the only solution up to $J^z/t\approx 4.25$. Transverse order, and with
it the $\Y$ order, appears only between $J^z/t\approx 6.25$ and $6.5$. By then the Ising moment is
$|\bm m_A|\approx 0.38$, three quarters of its classical value. Along the route of the main text, $J^z/t=2$
with $U$ increasing, transverse order appears between $U/t\approx 8.25$ and $8.5$, at $|\bm m_A|\approx 0.29$.
Fig.~\ref{fig:noU}(a) compares the two routes.
On the mean-field level, the $U$ term drives the longitudinal and transverse channels almost equally, and the two onsets of orders lie close together. Without it the transverse channel is driven by $J^\perp=0.3J^z$ alone and orders only
once the Ising moments are close to saturation. The $\Y$ state at $U=0$ is thus nearly classical.

\subsection{Collective modes}

\begin{figure*}[t]
\centering
\includegraphics[width=\textwidth]{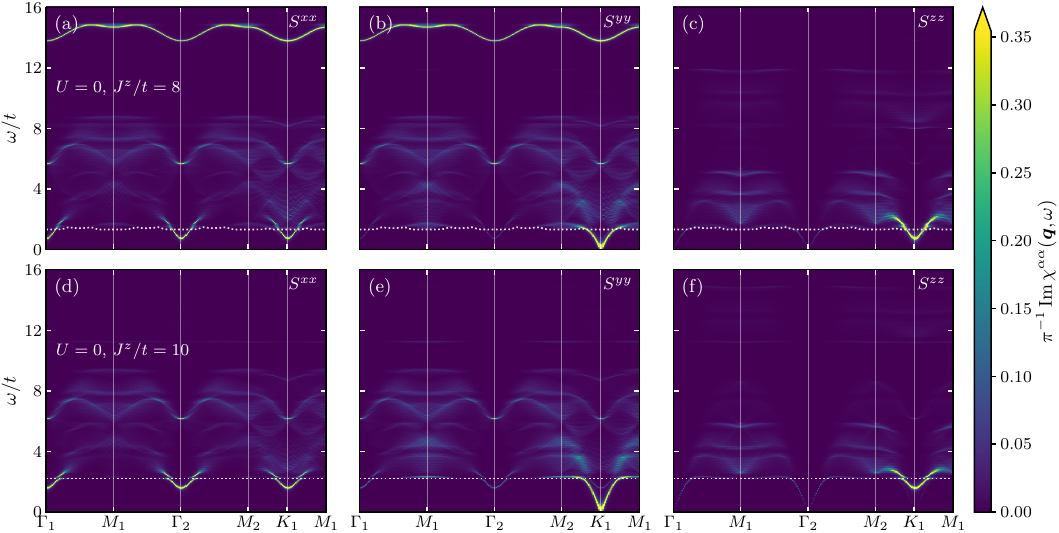}
\caption{Dynamical structure factor $S^{\alpha\alpha}(\bm q,\omega)$ of the $\Y$ state with $U=0$, at $J^\perp/J^z=0.3$, along the path of Fig.~\ref{fig:model}(b).
(a)-(c) $J^z/t=8$. (d)-(f) $J^z/t=10$. Columns show $\alpha=x,y,z$. The dotted line is the lower
edge of the two-spinon continuum.}
\label{fig:noUspec}
\end{figure*}

Fig.~\ref{fig:noUspec} shows the RPA corrected excitation spectra of $\Y$ at $J^z/t=8$ and $10$. The spectra differ
qualitatively from those of Fig.~\ref{fig:bare}. The continuum now starts far below the exchange
scale. The
collective modes, whose energy scale is set by the exchange, therefore lie inside the continuum over most
of the zone.
At the $M$ point neither point has a collective mode below the continuum, and $S^{\alpha\alpha}(\bm M,\omega)$ is a broad continuum without
sharp peaks (Fig.~\ref{fig:noU}(b)).
Therefore, in this phenomenological model, the exchange-only case
fails to produce the sharp mode at $M$, even though the gapless Goldstone mode is successfully captured.

\section{Effects of tuning \texorpdfstring{$U$}{U}}\label{app:uj}

\begin{figure*}[t]
\centering
\includegraphics[width=\textwidth]{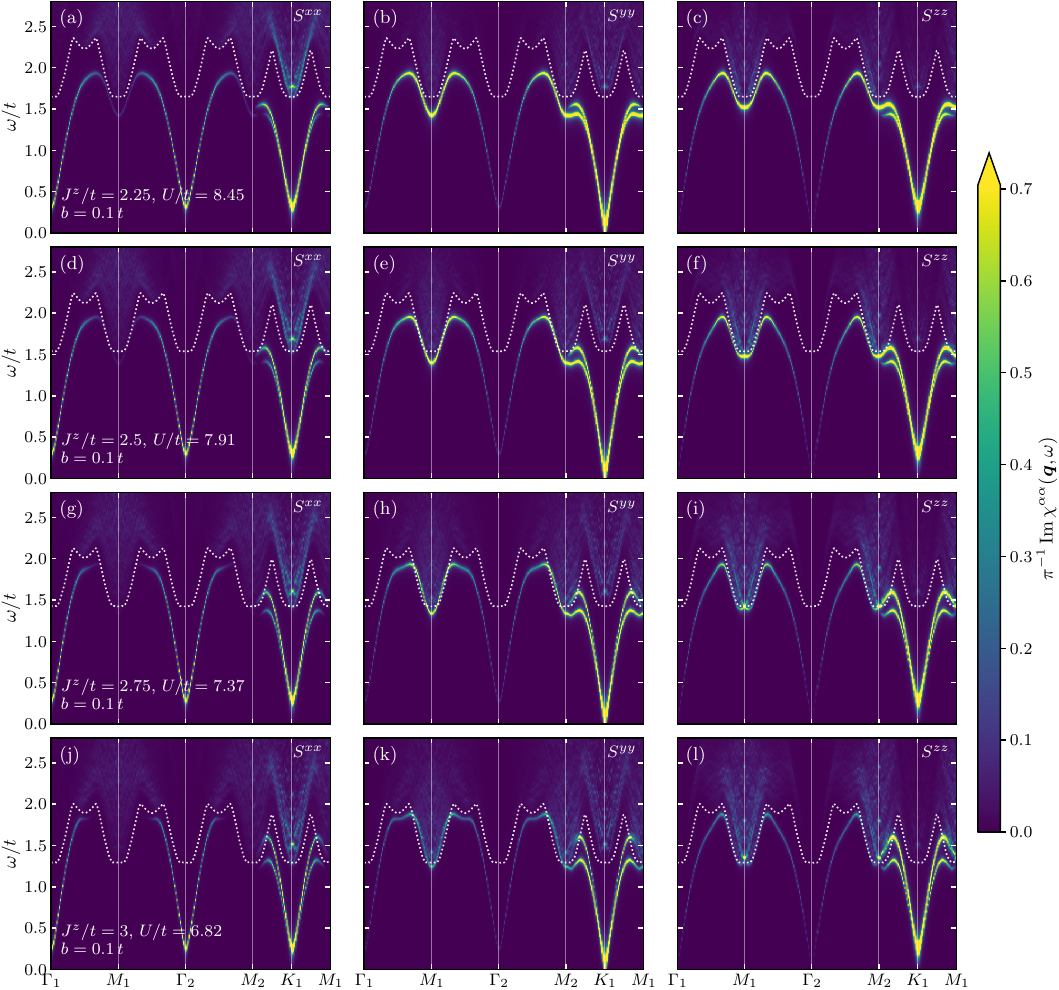}
\caption{Dynamical structure factor $S^{\alpha\alpha}(\bm q,\omega)$ of the $\Y$ state with
$|\bm m_A|=0.334$ where $U$ is traded for $J^z$, at $J^\perp/J^z=0.3$, stabilized using $b=0.1t$, along the path of Fig.~\ref{fig:model}(b). (a)-(c) $J^z/t=2.25$, $U/t=8.45$.
(d)-(f) $J^z/t=2.5$, $U/t=7.91$. (g)-(i) $J^z/t=2.75$, $U/t=7.37$. (j)-(l) $J^z/t=3$,
$U/t=6.82$.}
\label{fig:ujspec}
\end{figure*}

Here we investigate the effects of tuning $U$ more carefully. For better comparison, at each $J^z/t$ we choose $U$ so that the
largest moment stays the same at $|\bm m_A|\approx 0.334$. Namely, when $U$ is decreased, then $J^z$ is increased correspondingly. The self-consistent $\Y$ state
exists up to $J^z/t\approx 4.5$, $U/t\approx 3.46$. At fixed $|\bm m_A|$ the other moments shrink. Between $J^z/t=2$ and
$J^z/t=3$, $|\bm m_{B,C}|$ falls from $0.288$ to $0.253$ and $m^\perp$ from $0.218$ to $0.170$, because
the easy-axis exchange drives the longitudinal channel much more strongly than the transverse one.
Fig.~\ref{fig:ujspec} shows the excitation spectra of $\Y$ at four points in parameter space. The main change is that the continuum comes down relative to the collective modes. The modes lose weight as they approach the continuum edge and eventually disappear into the continuum.

\section{Effects of tuning anisotropy}\label{app:aniso}
\begin{table}[h]
\caption{Parameters of the stabilized $\Y$ state for (a) the uniform field $h=b$ of Eq.~\eqref{eq:b} and
(b) the moment-aligned field $h=\mu$ of Eq.~\eqref{eq:mu}, at $J^z/t=2$, $J_2=J_3=0$ and $U=U_c+0.636\,t$. $U_c$ denotes the critical value for the transverse onset of magnetic order, $h_c$ the field at which the tilt eigenvalue
at $\bm K$ changes sign (i.e. $\Y$ becomes stable), and $h$ the field used, all in units of $t$. The last row of each block contains the parameters used in the main text.}
\label{tab:parameters}
\begin{ruledtabular}
\begin{tabular}{d dd dd}
 \multicolumn{1}{c}{$J^\perp/J^z$} & \multicolumn{1}{c}{$U_c$} & \multicolumn{1}{c}{$U$}
 & \multicolumn{1}{c}{$h_c$} & \multicolumn{1}{c}{$h$}\\
\hline
\multicolumn{5}{l}{(a) Uniform field, $h=b$}\\
0.10 & 8.726 & 9.362 & 0.112 & 0.18\\
0.15 & 8.649 & 9.285 & 0.095 & 0.16\\
0.20 & 8.561 & 9.197 & 0.080 & 0.12\\
0.25 & 8.465 & 9.101 & 0.067 & 0.10\\
0.30 & 8.364 & 9.000 & 0.055 & 0.10\\
\hline
\multicolumn{5}{l}{(b) Moment-aligned field, $h=\mu$}\\
0.10 & 8.726 & 9.362 & 0.223 & 0.350\\
0.15 & 8.649 & 9.285 & 0.180 & 0.275\\
0.20 & 8.561 & 9.197 & 0.143 & 0.225\\
0.25 & 8.465 & 9.101 & 0.112 & 0.175\\
0.30 & 8.364 & 9.000 & 0.086 & 0.150\\
\end{tabular}
\end{ruledtabular}
\end{table}

\begin{figure}
    \centering
    \includegraphics[width=0.95\linewidth]{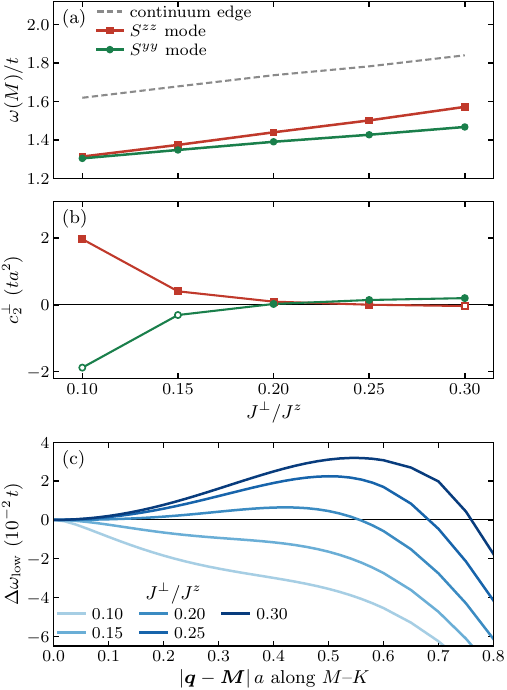}
    \caption{$\Y$ state stabilized with the $\mu$-field (see Sec.~\ref{sec:mu}) under changing anisotropy. (a) The collective modes at the $M$ point, with the dashed line indicating the continuum edge. (b) The curvature of the modes at the $M$ point. The open markers indicate negative curvature, i.e. no more roton minimum. (c) The dispersions of the collective modes along the $M$--$K$ direction, with $\Delta\omega_\mathrm{low}\equiv \omega_\mathrm{low}(\bm q)-\omega_\mathrm{low}(\bm M)$. The parameters used are listed in Table~\ref{tab:parameters}.}
    \label{fig:aniso-mu}
\end{figure}

\begin{figure}
    \centering
    \includegraphics[width=0.95\linewidth]{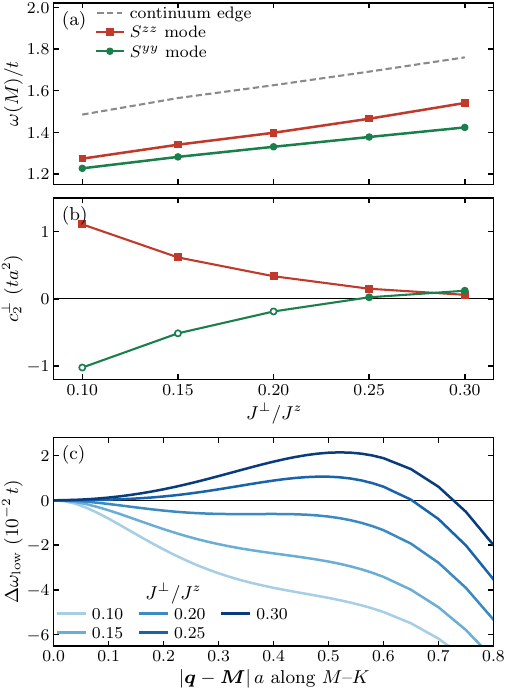}
    \caption{$\Y$ state stabilized with the $b$-field (see Sec.~\ref{sec:b}) under changing anisotropy, similar to Fig.~\ref{fig:aniso-mu}.}
    \label{fig:aniso-b}
\end{figure}
In this section, we study in more detail the effects of tuning the anisotropy $J^\perp/J^z$ in the exchange couplings, in particular on the roton mode at the $M$ point. Since the dip is robust along the $\Gamma_1$--$M_1$--$\Gamma_2$ path, we focus on the perpendicular path $M_1$--$K_1$ (see Fig.~\ref{fig:model}(b)). Whether the collective mode has a local minimum or not can be diagnosed by the curvature of the dispersion, which we define here using finite difference:
\begin{equation}
    c_2=\lim_{\delta q\to0}\frac{\omega(\bm M+\delta q\,\hat{\bm e})+\omega(\bm M-\delta q\,\hat{\bm e})-2\,\omega(\bm M)}{2\,\delta q^2},
\label{eq:curvature}
\end{equation}
with $\hat{\bm e}$ along $\Gamma$--$M$ for $c_2^\parallel$ and along $M$--$K$ for $c_2^\perp$.

Fig.~\ref{fig:aniso-mu} shows the results for the $\Y$ state stabilized using the moment-aligned $\mu$-field discussed in Sec.~\ref{sec:mu}. For each anisotropy, the required field strength for the stabilization is different, and the critical $U_c$ for the onset of the the transverse magnetic order is also different. We pick the parameters such that the mean-field state is not too close to the critical point and is well stabilized, shown in Table~\ref{tab:parameters}. Similarly, Fig.~\ref{fig:aniso-b} shows the case where the $\Y$ state is stabilized with a uniform $b$-field along the easy-axis direction. In both cases, the curvature along the $M$--$K$ direction at the $M$ point becomes negative when the anisotropy is sufficiently large. Namely, the local roton minimum at the $M$ point disappears at strong anisotropy.

\bibliography{refs}
\end{document}